\long\def\UN#1{$\underline{{\vphantom{\hbox{#1}}}\smash{\hbox{#1}}}$}
\def\NP{\vfil\eject}
\def\NI{\noindent}
\def\NL{\hfill\break}
\magnification=\magstep 2
\overfullrule=0pt
\hfuzz=16pt
\voffset=0.0 true in
\vsize=8.8 true in
\baselineskip 20pt
\parskip 6pt
\hoffset=0.1 true in
\hsize=6.3 true in
\nopagenumbers
\pageno=1
\footline={\hfil -- {\folio} -- \hfil}

\ 

\ 
 
\centerline{{\bf Diffusional Growth of Colloids}}
 
\ 

\centerline{Dima Mozyrsky and Vladimir Privman$^*$}

\ 
 
\centerline{\sl Department of Physics, Clarkson University}
\centerline{\sl Potsdam, New York 13699--5820, USA}

\vfill

\centerline{\bf Abstract} 

\ 

We consider incorporation of particle detachment in 
Smoluchowski model of colloidal growth.
Two approaches are considered, utilizing phenomenological 
rate equation and exact large-time results.
Our main conclusion is that the value of the large-time 
diffusing particle concentration at the aggregate
surface is the only parameter needed to describe the 
added effect of detachment. Explicit expression is given 
for the particle intake rate.

\vfill

\NI{}$^*$Electronic mail: privman@clarkson.edu

\NP

\NI{\bf 1.\ \ Introduction}

\ 
 
In this work we address aspects of the theory of growth of 
aggregates by capture of small 
diffusing particles (singlets), such as atoms, molecules, 
etc., with focus on processes relevant in colloidal growth. The 
celebrated work of Smoluchowski [1], reviewed, e.g., 
in [2], considered the process of diffusional 
capture of particles as follows. The growing aggregate 
is modeled ``adiabatically'' as a sphere 
of radius $R$. The concentration of particles is 
initially $c_p$ ($p$ stands for particles). 
Next, the radially 
symmetric concentration of particles at time 
$t$, $c(r,t)$, is calculated by solving the 
diffusion equation with the absorbing boundary 
condition at $r=R$, i.e., $c(R,t>0)=0 $. The resulting 
particle intake 
$\Phi$ (number per unit time) into the sphere is 
usually taken as the large-time (steady-state) 
limit of the following expression obtained from this solution:

$$ \Phi (t) = 4 \pi D R c_p \left( 1 + { R \over 
\sqrt{ \pi D t } } \right) \; , \eqno(1.1) $$

\NI where $D$ is the singlet diffusion constant. The 
value of $\Phi (\infty)$ is then used to estimate 
the growth rate of the spherical aggregate.

In colloidal growth, a popular model has been that 
of ``burst nucleation'' by LaMer and Dinegar 
[3-4]. In this model, one adopts concepts from the 
classical nucleation theory to colloidal aggregation. 
Specifically, for sufficiently small aggregate radii 
$R$, it is assumed that the aggregate 
sizes are
distributed according to the equilibrium 
condition with the Boltzmann factor determined 
by the aggregate free energy $G(R)$. This equilibrium 
distribution is usually assumed to apply for $R<R_c$, 
where the critical radius $R_c$ corresponds to the 
maximum of the free energy.

Larger aggregates, those with $R>R_c$, are assumed to 
grow irreversibly by the diffusional mechanism
(1.1) and possibly by cluster-cluster aggregation, 
depending on the precise nature of forces between 
aggregates. The whole picture is ``adiabatic'' in 
that many of the quantities involved, specifically, 
$G$, $R_c$, $c_p$, evolve in time, presumably on the 
time scales much larger than the times needed to establish 
the asymptotic large-time diffusional flux; see (1.1). 
Thus, for the diffusion part of the description, 
the steady state results nearly always suffice.

For clusters with $R<R_c$, there must be dynamical 
mechanisms that lead to equilibration, if the 
``nucleation'' droplet-model concept is indeed 
applicable to colloids. In addition to the 
mechanisms that correspond to internal rearrangement 
and restructuring, there must be also
\UN{detachment} of particles. Indeed, the aggregates 
cannot only grow, by capturing singlets, while at
the same time achieve equilibrium distribution. Since 
the critical cluster size $R_c$ can be actually quite 
large, it follows ``by continuity'' that detachment 
will also occur for clusters  with $R>R_c$ which are 
not in equilibrium.

The main goal of the present work is to incorporate 
detachment process in Smoluchowski model. Thus, we 
focus on the diffusion problem. We assume that 
the aggregates are large, $R>R_c$, and that their 
radii vary slowly with time so that the time scales 
of singlet diffusional processes are fast in comparison. 
Furthermore, by focusing on the diffusion problem we 
avoid discussion of cluster-cluster aggregation and 
of detachment of multiplets, as well as of internal 
restructuring processes. The latter are present not 
only for small clusters but also for large clusters 
because without internal restructuring, diffusion-limited 
colloidal growth would result in fractals, e.g. [5]. Under
properly chosen experimental conditions, e.g. [6] and
references therein, the 
aggregates are as dense
as the bulk material and have definite shapes, 
frequently spherical as assumed here, and 
semicrystalline structure.

Our second goal is to emphasize a difference 
between the colloidal growth and classical nucleation.
Specifically, we note that the intake expression 
(1.1) for $t=\infty$ is proportional to $R$ rather 
than to the
surface area of the collector aggregate $4 \pi R^2$. 
In ordinary metastable-to-stable two-phase nucleation, 
when the transport mechanisms are not diffusional, 
for large clusters one can assume that each unit of 
the surface area captures and emits equal amounts 
of atoms or molecules. The net intake rate is then 
proportional to the surface area [7,8]. We would 
like to explore failure of this ``surface 
extensivity'' argument for colloids and include 
detachment in this discussion. A related matter 
is the mere existence of the ``surface layer'' since 
intuitively it seems to be connected with the
``per-unit-area'' adsorption and detachment.

Finally, we would like to have few-parameter 
approaches that retain the simplicity and exact 
solvability of Smoluchowski model.
We note that numerous modifications of 
Smoluchowski model have been proposed in the literature; 
see [2,9] for reviews. The most studied case in the field 
of colloids has been deposition of particles in a 
low- or intermediate-density
monolayer on a solid substrate, without incorporation 
of the deposited particle into the material of the 
collector by internal restructuring. Detachment from 
such monolayers is approximately proportional to the 
amount of matter deposited. It is modeled by 
phenomenological rate equations for the surface 
coverage coupled with expressions for the diffusional 
or convective-diffusional flux [10,11]. A detailed 
mathematical study of one such boundary condition 
was published by Agmon [12]. Experimental results 
have been reviewed in [11,13].

Another application of Smoluchowski approach has 
been in the field of chemical reaction rates
in the diffusion-limited regime, reviewed in [9]. 
Here $R$ is no longer large. Instead of detachment 
one has partial reaction, i.e., some particles may 
not react on encountering a collector. This is 
mathematically described by the radiation boundary 
condition [14].

The outline of the rest of this article is as follows. 
In Section~2, we consider a phenomenological model 
for the boundary condition at the aggregate surface 
that allows for detachment in Smoluchowski model. In 
Section~3,
we derive a generalized radiation boundary condition 
which is exact in the large-time limit. The diffusion 
problem with the latter boundary condition is solved 
in Section~4. Finally, in Section~5 we
compare the two approaches and conclude that for all 
practical purposes one can actually use a simpler 
boundary condition which only involves one new 
parameter: the concentration of the diffusing 
particles right at the aggregate surface. 

\NP 

\NI{\bf 2.\ \ Phenomenological Rate Equation Model}

\ 

In the original Smoluchowski model [1], the radially 
symmetric version of the diffusion equation,

$$ {\partial c \over \partial t} = r^{-2} {\partial 
\over \partial r} \left( {r^2} D
{\partial c \over \partial r} \right) \eqno(2.1) $$

\NI is solved for $r \geq R$, with the initial condition

$$ c(r,t=0) = c_p \eqno(2.2) $$

\NI and boundary condition

$$ c(r=R,t>0)=0 \; . \eqno(2.3) $$

\NI The discontinuity in the value of the concentration 
at $r=R$ (at $t=0$ as compared to that at $t>0$),
shared by many modifications of the problem,
leads to singular behavior at $t=0$, see (1.1), that 
has been of some concern in the literature on 
chemical reaction applications [2,9].

One can consider various modifications of the 
relations (2.1)-(2.3) to improve the model and
incorporate effects other than the irreversible 
adsorption at $r=R$ expressed by (2.3). In this 
section we consider a phenomenological 
modification of the boundary condition (2.3) 
to allow for detachment. We
use Langmuir-type mean-field rate equation approach 
popular in applications of this sort in colloid 
literature, e.g. [15]. Thus, we propose the relation

$$ {\partial c \over \partial t}=-Kc+k \; , 
\;\;\;\;\; {\rm at}\;\; r=R \; , \eqno(2.4)$$

\NI to replace (2.3). Here we assume that diffusing 
particles that reach the surface are incorporated 
in the aggregate structure at the rate $Kc$ 
proportional to their concentration
at $R$. The second term corresponds to detachment, 
and we assume that this rate only depends on 
the internal processes so there is no dependence 
on the external diffuser concentration. 

In order to solve (2.1) with (2.2), (2.4), we 
define Laplace Transform of the concentration,

$$ n(r,s)= \int_0^\infty dt \, e^{-st} c(r,t) \; , 
\eqno(2.5) $$

\NI which satisfies the equation

$$ s n - c_p = r^{-2} {\partial \over \partial r} 
\left( {r^2} D
{\partial n \over \partial r} \right) \; . \eqno(2.6) $$

\NI Here we already used the initial condition. The 
general solution of this equation is given by

$$ n={c_p \over s} + {A(s)\over r}e^{-\sqrt{s/D}\, 
r}+{B(s)\over r}e^{\sqrt{s/D}\, r} \; , \eqno(2.7) $$

\NI where we set $B=0$ to have a bounded solution 
as $r \to \infty$. The boundary condition at $R$
becomes

$$s n - c_p = -Kn+ks^{-1} \; , \;\;\;\;\; {\rm at}\;\; 
r=R \; . \eqno(2.8)$$

\NI The final result for the concentration is

$$ n={c_p \over s} - \left({R \over r}\right) 
{c_pK-k \over s(s+K)}e^{-\sqrt{s/D}\, (r-R)} \; , 
\eqno(2.9) $$

\NI The full time-dependence of the concentration 
can be expressed in terms of error functions, etc., 
by using the tables of Inverse Laplace Transform. 
However, the result is quite cumbersome and 
unilluminating. We present a simpler limiting 
expression shortly.

For Laplace Transform of the intake rate 
$\Phi(t)$, denoted $\phi(s)$, we get

$$ \phi=4 \pi D R{c_pK-k \over s(s+K)}
\left(1 + R \sqrt{s/D} \right) \; , \eqno(2.10) $$

\NI where we used

$$ \Phi(t)=4\pi R^2 D \left( {\partial c \over \partial r} 
\right)_{r=R} \; . \eqno(2.11) $$

\NI The time-dependence of the intake rate is then

$$\Phi(t)=4\pi D R \left(c_p K -k\right) 
\left({1-e^{-Kt} \over K} +
{2R \over \sqrt{\pi D K} } e^{-Kt} 
\int_0^{\sqrt{KT}} dx \, e^{x^2}\right) \, . \eqno(2.12) $$

We note that this rate expression is actually 
regular for small times. However, for colloid 
applications we
are interested in large times. For $t \gg 1/K$, 
the limiting form of (2.12) is quite similar 
to Smoluchowski result (1.1). Keeping 
only the leading time-dependent term, we have

$$ \Phi (t) \simeq 4 \pi D R \left(c_p - {k \over K} 
\right) \left( 1 + { R \over \sqrt{ \pi D t } } \right) 
\; . \eqno(2.13) $$

\NI The only change is the reduction of the rate 
due to detachment, proportional to the
ratio $k/K$ (the intake rate can actually become 
negative if the detachment is fast enough).

We now turn to the concentration and present its 
leading time dependence in the limit $t \gg 1/K$,

$$ c(r,t) \simeq c_p -\left(R\over r\right)
\left(c_p - {k \over K} \right) {\rm Erfc}
\left({r-R \over 2 \sqrt{  D t }}\right) 
\; . \eqno(2.14) $$

\NI As $t \to \infty$, the concentration profile 
develops a long-range tail $\sim R/r$ that has no 
length scale of order 1 associated with it. In fact, 
the perturbation of the diffuser particle 
concentration extends for distances comparable to $R$. 
The vanishing of the term $\sim R^2$ in (2.13),
so that only the ``nonextensive'' contribution $\sim R$ 
survives as $t\to \infty$, is associated with this 
property. Obviously, it is dependent on the transport 
mechanism being diffusional. Note that the terms 
retained in (2.13), (2.14) vary on the time scales set by
$R^2/D$. Since we are interested in large $R$ values, 
this quantity is much larger than the ``transient'' 
time scale $1/K$ present in the exact expressions.
 
\NP 

\NI{\bf 3.\ \ Model with Boundary Layer}

\ 

In this section we address the question of 
whether the boundary
condition at $r=R$ can be derived from a 
diffusional model. We will
keep the assumption that the diffusion is 
free for $r>R$ in the sense that
there is no external potential. However, let 
us assume that in a certain
boundary region from $R-\ell$ to $R$, where 
$\ell \ll R$, the particles
move in the potential energy $U(r)$ owning to 
their interaction with
the aggregate. As $r \to R-\ell$ from above, 
the diffusional picture breaks down as the 
particles are subject to the internal 
restructuring mechanisms in the aggregate. 
We will assume that this dense aggregate 
structure can be represented by the boundary condition

$$ c(R-l,t)=c_0 \; . \eqno(3.1) $$

\NI In fact, the diffusion constant $D$ will be 
affected by the increased particle concentration, 
proximity to the semicrystalline aggregate, and
external potential which itself can be partly 
functionally $c$-dependent.
We ignore all these complications and keep $D$ 
constant.

Thus, our model is presented in Figure~1. We 
assume that $c_0 \gg c_p$ is
a fixed quantity that only depends on the 
aggregate material. We also put

$$ V(r) = U(r) /k_B T \; \eqno(3.2) $$

\NI and assume that $V(r)$ is a fixed function 
of the difference $r-R$, vanishing for $r \geq R$. 
This potential energy (in units of $k_B T$) may 
represent a truncated version of the full 
interaction energy at $r$. For ``free'' 
diffusional aggregation, we can assume 
that the interactions for $r>R$ are weak. 
However, in the boundary layer of size $\ell$, large 
negative energy is
expected, as sketched in Figure~1. Thus, we denote 

$$ V(R-\ell)\equiv -V_0 , \eqno(3.3) $$

\NI with $V_0 \gg 1$. Furthermore, we expect 
that $|V(r)| \gg 1$ in most
of the boundary layer.

The diffusion equation is replaced by one with 
the potential energy term:

$$ {\partial c \over \partial t} = r^{-2} 
{\partial \over \partial r} \left[ {r^2} D \left(
{\partial c \over \partial r} + c \, {d V 
\over d r} \right) \right] \; . \eqno(3.4) $$

\NI We now offer exact considerations which 
are valid only in the steady-state $t\to \infty$ 
limit. Then the diffusion equation can be integrated 
as follows:

$$ D \left(
{d c \over d r} + c \, {d V \over d r} \right) = 
{\Phi \over 4 \pi r^2} \; . \eqno(3.5) $$

\NI Here $c(r)$ denotes $c(r,\infty)$, and $\Phi=\Phi 
(\infty)$ is the (constant) intake rate into 
a sphere of radius $r$ at time $t=\infty$.

Now the potential $V(r)$ is continuous but its 
derivative is likely discontinuous; see Figure~1.
It then follows from (3.5) that the concentration 
$c(r)$ is continuous but the derivative $dc/dr$ 
is not. Consider the limit $r\to R$ from above, 
denoted by + here. We have the following expression 
involving the derivative:

$$ \Phi = 4 \pi R^2D\left({dc \over dr}\right)_{r\to 
R+} \; . \eqno(3.6) $$

\NI Next, we solve the steady-state diffusion 
equation for $r<R$ for the concentration $c(r)$ 
by direct integration and use of the boundary 
values (3.1) and (3.3). We also use $V(R)=0$ to get, for the
the concentration at $R$,

$$ c(R)= c_0 e^{-V_0} + {\Phi \over 4 \pi D}\int_{R-\ell}^R 
dr \, {e^{V(r)}\over r^2}
 \; . \eqno(3.7) $$

\NI Finally, we utilize (3.6) to obtain the following relation:

$$ c(R)=\alpha \left({dc \over dr}\right)_{r\to R+} 
+ \beta \; , \eqno(3.8) $$

\NI where

$$ \alpha=R^2 \int_{R-\ell}^R dr \, {e^{V(r)}\over r^2}
 \; , \eqno(3.9) $$

$$ \beta=c_0 e^{-V_0} \; . \eqno(3.10) $$

Since relation (3.8) applies with the derivative 
at $R+$, it can be used as the boundary condition 
for the free diffusion at $r \geq R$. We can thus 
lump all the effects of the boundary layer in two 
parameters $\alpha$ and $\beta$. Such boundary 
condition resembles the radiation boundary condition 
used in the chemical reaction applications [9,14], 
which assumes that the concentration is proportional 
to its derivative. However, there are two important 
differences. Firstly, there is the additional constant 
term $\beta$. Secondly, the coefficient $\alpha$ in 
our case is only weakly dependent on $R$. The integral 
in (3.9) is obviously of order

$$\alpha \simeq \ell e^{-V_0} \; . \eqno(3.11) $$

\NI Thus, the main result of this section is that 
the problem with the boundary layer can be replaced, 
at least for large times, by an equivalent problem 
of free diffusion at $r \geq R$, as long as we use 
our generalized radiation boundary condition at $R$, 

$$ c=\alpha \,{\partial c \over \partial r} + \beta  
\; , \;\;\;\;\; {\rm at}\;\; r=R \; . \eqno(3.12) $$

\NP 

\NI{\bf 4.\ \ Generalized Radiation Boundary Condition}

\  

Before comparing, in the next section, the two choices of the boundary 
condition: phenomenological (2.4) and radiation 
(3.12), let us work out, in this section, what would be 
the time 
dependence of the free diffusion problem at $r \geq 
R$ if the boundary condition (3.12) were used 
for all times. We go through 
steps similar to Section~2 and only emphasize the final 
expressions. In Laplace Transform version, the 
boundary condition becomes

$$ n=\alpha \, {\partial n \over \partial r} + 
\beta s^{-1} \; , \;\;\;\;\; {\rm at}\;\; r=R \; . \eqno(4.1) $$

\NI For the concentration and intake rate, we get

$$ n={c_p \over s} - \left({R \over r}\right) 
{c_p-\beta \over s \left(1+\alpha R^{-1} + \alpha  
\sqrt{s/D}\right)}e^{-\sqrt{s/D}\, (r-R)} \; , \eqno(4.2) $$

$$ \phi=4 \pi D R\, {c_p-\beta \over s \left(1+\alpha R^{-1} + 
\alpha  \sqrt{s/D} \right) }
\left( 1 + R \sqrt{s/D} \right) \; . \eqno(4.3) $$

These results can be inverted to get the time 
dependence since they only involve tabulated functions.
For instance,

$$ \Phi (t) = 4 \pi DR(c_p-\beta){R\over 
\alpha+R}\left[1+{R\over \alpha}\, e^{(\alpha+R)^2 
\alpha^{-2}
R^{-2} D t } \, {\rm Erfc} \left({\alpha+R \over 
\alpha R} \sqrt{Dt} \right) \right] \; . \eqno(4.4) $$

\NI This expression is well-behaved at $t=0$. As 
in Section~2, though, it is more useful to analyze 
limiting results for large $R$ and large times.
We note that $\beta$ is the product of a large 
quantity ($c_0 \gg c_p$) and a small Boltzmann factor
($e^{-V_0} \ll 1$); see (3.10). However, $\alpha$ 
is the product of two small quantities: $\ell \ll R$ and
the Boltzmann factor; see (3.11). Therefore, we can 
always assume that $R \gg \alpha$. Given this assumption, 
one then identifies the transient time scale via the condition

$$ t \gg \alpha^2 / D \; . \eqno(4.5) $$

\NI The asymptotic expression for the intake rate becomes

$$ \Phi (t) \simeq 4 \pi D R \left(c_p - \beta \right) 
\left( 1 + { R \over \sqrt{ \pi D t } } \right) \; , \eqno(4.6) $$

\NI which is the same as (2.13) provided we identify

$$ \beta = {k \over K} \; . \eqno(4.7) $$

\NI Similarly, with (4.7), the concentration is given by (2.14).

\NP

\NI{\bf 5.\ \ Constant Boundary Condition. Discussion}

\ 

In view of the above discussion it is tempting to 
combine the phenomenological model which was introduced 
for all $t$-values with the exact expression for the 
boundary condition at $t=\infty$. We could write something like 

$$  K^{-1} {\partial c \over \partial t} = -c+\beta + 
\alpha \,{\partial c \over \partial r} \; , \;\;\;\;\; 
{\rm at}\;\; r=R \; . \eqno(5.1) $$

\NI Here $\beta$ is given by (3.10), $\alpha$ by (3.9) 
and (3.11), and it is also tempting to assume that there 
is only one transient time scale so that

$$ K \propto D/\alpha^2 \; . \eqno(5.2) $$

However, as long as we are interested only in the 
large $R$ and large time behavior, for all practical 
purposes we can use instead the boundary condition

$$ c=\beta  \; , \;\;\;\;\; {\rm at}\;\; r=R \; . \eqno(5.3) $$

\NI Indeed, the value of the concentration at $R$ 
for large times, which is exactly $\beta=k/K$, 
independent of $R$, in both models, see (2.9) 
and (4.2), is the only parameter needed to calculate 
the modification of the
asymptotic intake rate due to detachment. With this 
boundary condition for all times, the results (2.13) 
or (4.6), and (2.14), become exact. The price paid 
is that the singularity at $t=0$ is back. However, 
only large $t$ results are of interest in colloidal 
growth applications.

In summary, a combination of phenomenological and exact 
considerations yields expressions for the diffusional 
intake rate of particles in Smoluchowski model with 
added detachment. The intake rate
is reduced (it can become negative) and its value can 
be calculated from the large-time concentration at the 
collector surface ($\beta$) and the value far away 
from the surface ($c_p$). The only added parameter 
is thus $\beta$. At large times, the concentration 
profile develops a slowly varying contribution $\sim R/r$ 
and the perturbation of the concentration due to the 
collector surface extends on length scales $\sim R$, 
far beyond the dimension of the ``boundary layer'' 
$\ell \ll R$. This added concentration variation is not 
present in classical nucleation models with nondiffusional 
transport. The main result is the
nonextensive ($\sim R$ instead of $\sim R^2$) 
behavior of the intake rate in the $t\to \infty$ limit
in the diffusional model.

\NP

\centerline{\bf References}{\frenchspacing

\ 

\item{1.} M. v. Smoluchowski, {\it Versuch 
einer mathematischen Theorie der 
Koagulationkinetik kolloider L\"osunger}, Z. 
Phys. Chem. {\bf 29}, 129 (1917).

\item{2.} G. H. Weiss, {\it Overview of 
Theoretical Models of Reaction Rates}, 
J. Statist. Phys. {\bf 42}, 3-36 (1986).

\item{3.} V. K. LaMer and R. H. Dinegar, {\it 
Theory, Production 
and Mechanism of Formation of Monodispersed 
Hydrosols}, J. Amer. Chem. 
Soc. {\bf 72}, 4847-4854 (1950).

\item{4.} V. K. LaMer, {\it Nucleation in Phase 
Transitions}, 
Ind. Eng. Chem. {\bf 44}, 1270-1277 (1952).

\item{5.} D. W. Schaefer, J. E. Martin, P. 
Wiltzius and
D. S. Cannell, {\it Fractal Geometry of Colloidal 
Aggregates}, Phys. Rev. Lett. {\bf 52}, 2371-2374 (1984).

\item{6.} V. Privman, D. V. Goia, J. Park and 
E. Matijevi\'c,
{\it Mechanism of Formation of Monodispersed 
Colloids by Aggregation 
of Nanosize Precursors}, preprint 
http://xxx.lanl.gov/abs/cond-mat/9809167 (1998).

\item{7.} J. E. McDonald, {\it Homogeneous Nucleation 
of Vapor Condensation. II. Kinetic Approach}, Am. J. 
Phys. {\bf 31}. 31-41 (1962).

\item{8.} K. F. Kelton and A. L. Greer, {\it Test of 
Classical Nucleation Theory
in a Condensed System}, Phys. Rev. B {\bf 38}, 
10089-10092 (1988).

\item{9.} R. M. Noyes, {\it Effects of Diffusion 
Rates on Chemical Kinetics},
Prog. React. Kin. {\bf 1}, 129-160 (1961).

\item{10.} S. L. Zimmer and B. E. Dahneke, {\it 
Resuspension of Particles: the Range of Validity 
of the Quasi-stationary Theories}, J. Coll. Interf. 
Sci. {\bf 54}, 329-338 (1976). 

\item{11.} N. Kallay, E. Barouch and E. Matijevi\'c, 
{\it Diffusional Detachment of Colloidal Particles 
from Solid/Solution Interfaces}, Adv. Coll. Interf. 
Sci. {\bf 27}, 1-42 (1987). 

\item{12.} N. Agmon, {\it Diffusion with Back 
Reaction}, J. Chem. Phys. {\bf 81}, 2811-2817 (1984).

\item{13.} E. Matijevi\'c, E. Barouch and N. 
Kallay, {\it Kinetics of Diffusional Detachment 
of Colloidal Particles from Surfaces}, Croat. 
Chem. Acta {\bf 60}, 411-428 (1987). 

\item{14.} F. C. Collins and G. E. Kimball, 
{\it Diffusion-Controlled Reaction Rates},
J. Coll. Sci. {\bf 4}, 425-437 (1949).

\item{15.} V. Privman, H. L. Frisch, 
N. Ryde and E. Matijevi\'c, {\it Particle 
Adhesion in Model Systems. Part 13. --- 
Theory of Multilayer Deposition}, J. Chem. Soc. 
Faraday Trans. {\bf 87}, 1371-1375 (1991).

}

\NP

\centerline{\bf Figure Caption}

\ 

\NI Figure 1:\ \ \ Schematic representation of the 
truncated dimensionless \NL
\phantom{Figure 1:\ \ \ }potential energy $V(r)$, 
defined in Section~3.

\bye